\documentclass[letterpaper, 10 pt, conference]{ieeeconf}

\IEEEoverridecommandlockouts                              

\usepackage{cite}
\usepackage{amsmath,amssymb,amsfonts}
\usepackage{amsthm}                 
\usepackage{float}
\newcommand{\field}[1]{\mathbb{#1}}
\newcommand{\R}{\field{R}}

\usepackage{graphicx}

\usepackage{xcolor}
\usepackage{array}
\usepackage{booktabs}
\definecolor{bluem}{HTML}{0053D6}
\usepackage{hyperref}
\usepackage[nameinlink,capitalize,noabbrev]{cleveref}

\DeclareMathOperator*{\argmin}{arg\,min}

\author{Turki Bin Mohaya$^\star$ \quad Maitham F. AL-Sunni$^\star$ \quad John M. Dolan \quad Peter Seiler%
\thanks{$^\star$ These authors contributed equally to the work.}%
\thanks{T. Bin Mohaya and P. Seiler are with the Department of Electrical Engineering and Computer Science at the University of Michigan, Ann Arbor, MI, USA. Email: {\tt \{turki,pseiler\}@umich.edu}}%
\thanks{M. F. AL-Sunni is with the Department of Electrical \& Computer Engineering, Carnegie Mellon University, Pittsburgh, PA,
USA. Email: {\tt malsunni@andrew.cmu.edu}}%
\thanks{J. M. Dolan is with the Robotics Institute, Carnegie Mellon University, Pittsburgh, PA, USA. Email: {\tt jdolan@andrew.cmu.edu}}%
}

\begin{document}
\title{\LARGE \bf Generalizable Optimal Control with Transformers: 
\\ One Policy Across Diverse Systems}

\maketitle

\begin{abstract}
Classical optimal control designs a separate controller for each plant. Even for the Linear Quadratic Regulator (LQR), every new model must be identified and its Riccati equation re-solved. We ask whether a single learned policy can instead serve an entire family of systems, and we show that one transformer can. We train the policy to imitate optimal LQR state feedback across a collection of heterogeneous Multiple-Input, Multiple-Output (MIMO) Linear Time-Invariant (LTI) systems that differ in their state and input dimensions and in their cost objectives. A shared representation lets the same parameters control every member of the family. It combines system-wise standardization, zero-padding and masking across dimensions, and an explicit encoding of the cost matrices. At run time the policy maps a short window of recent states and the specified cost to a control action. It uses no plant matrices and identifies the dynamics implicitly from the state history. We evaluate on $28$ simulated systems over {$9{,}675$} closed-loop rollouts, and no unstable rollout was observed in any of them. On the systems seen during training it attains a median relative sub-optimality of $0.022\%$, even under parameter perturbations of up to {$\pm10\%$}. It transfers to unseen systems with lightweight fine-tuning, reaching a median sub-optimality of $0.19\%$. These results support transformers as generalizable near-optimal controllers for structured families of linear systems.

\end{abstract}


\section{Introduction}
\label{sec:introduction}
In robotics and cyber-physical systems, the plant model is rarely fixed: operating-point shifts, payload changes, contact conditions, temperature effects, and wear mean the controller faces not one plant but a family of plausible models. Classical optimal control designs each member separately---the discrete-time LQR~\cite{anderson1990optimal} requires identifying each new model before re-solving its Riccati equation---gating redesign on fresh data as much as on computation. We ask: can a single transformer policy be trained to produce near-optimal feedback actions across an entire family of linear systems?

\begin{figure}[t]
    \centering
    \includegraphics[width=\linewidth]{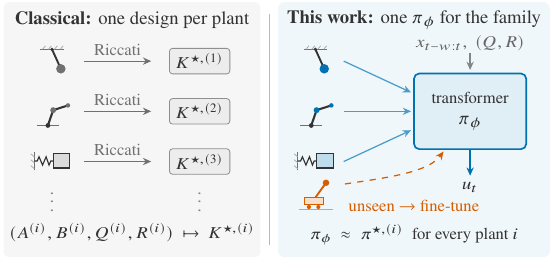}
    \caption{{Unlike} the classical LQR approach, we train a single transformer policy $\pi_\phi$ that maps a window of recent states and the cost matrices $(Q,R)$ to a near-optimal control $u_t$ for every member of the family, with unseen systems handled by {lightweight} fine-tuning.}
    \label{fig:overview}
\end{figure}

The premise is that a short window of closed-loop states reveals the generating dynamics, so a sequence model can identify the system implicitly and act accordingly. Sequence modeling has long relied on recurrent architectures~\cite{hochreiter1997long}, whose backpropagation-through-time training suffers from vanishing and exploding gradients~\cite{pascanu2013difficulty}. Transformers~\cite{vaswani2017attention} dispense with recurrence: self-attention lets all elements of the input history interact directly and in parallel---decisive when training on millions of short windows from many systems---making them well suited to history-dependent feedback control. We cast cross-system optimal control as sequence-to-control learning, training a single transformer policy on optimal LQR state-feedback rollouts over many model variants and initial conditions, across multiple cost settings{; \autoref{fig:overview} illustrates this {setup}}.

\subsection{Related work}
\label{sec:related_work}
Classical robust and gain-scheduled methods extend the per-plant Riccati solution to bounded uncertainty~\cite{zhou1996robust,shamma1988gain}, but still yield one controller per plant or uncertainty set. A complementary line learns LQR gains by policy optimization, where policy gradient converges globally~\cite{fazel2018global} (see~\cite{hu2023toward} for a survey). Extensions reach plant families through MAML-style initializations~\cite{toso2024meta}, federated learning over heterogeneous systems~\cite{wang2023federated}, a common gain shared across tasks with {sub-optimality} and stability bounds~\cite{stamouli2025policy,toso2026multitask}, and multi-task imitation of linear controllers~\cite{zhang2023multi}. In all of these, the learned object remains a per-task gain, a shared initialization, or a compromise gain whose {sub-optimality} grows with task heterogeneity, always at a fixed state dimension.

Generalist neural policies pursue breadth rather than per-instance optimality. Neural approximation of feedback laws is classical~\cite{lewis1999neural}, and reinforcement learning trains universal policies across model distributions via online system identification and dynamics randomization~\cite{yu2017preparing,peng2018sim}. At scale, cross-embodiment transformers~\cite{brohan2023rt1,Doshi24-crossformer} and low-level foundation models for dynamics and control~\cite{xiao2024anycaranywherelearninguniversal,xu2026car,eschmann2025raptor} generalize across platforms. None of these distills optimal-control supervision into one policy for a family of systems.

Transformers, as universal sequence-to-sequence approximators~\cite{yun2019are}, are a natural architecture for this setting. They have been used to model physical systems~\cite{geneva2022transformers}, to accelerate trajectory optimization and MPC~\cite{wang2025quattro,celestini2024transformer}, and to cast control as sequence modeling~\cite{chen2021decision}. Closest to us, one transformer trained across random LTI systems learns optimal filtering in-context~\cite{du2023can}, and in-context learning yields class models for system identification~\cite{forgione2023from}. Both address estimation and identification rather than closed-loop control. We target the complementary closed-loop control problem: our policy conditions on recent states and the cost, so that one set of weights approaches each instance's own optimum across plant families and state dimensions.

\subsection{Contribution and Organization}
The contributions of this paper are {threefold}. First, we formulate cross-system LQR imitation over heterogeneous Linear Time-Invariant (LTI) systems with differing state and input dimensions. Second, we introduce a shared representation and masking scheme, together with explicit conditioning on the LQR cost matrices, that lets one policy train across different state and input dimensions and across different cost specifications. Third, we demonstrate in simulation that the policy is near-optimal and robust over a broad family of perturbed systems, and adapts rapidly to unseen systems via few-shot fine-tuning.

The rest of the paper is organized as follows. Section~\ref{sec:problem_formulation} states the cross-system LQR problem and the universal policy we seek. Section~\ref{sec:appraoch} presents the approach, including the shared representation, the transformer architecture, training, and fine-tuning for unseen systems. Section~\ref{sec:results} evaluates the policy across the system family. Section~\ref{sec:conclusions} concludes.

\section{Problem Formulation}
\label{sec:problem_formulation}

Consider the LTI system
\normalsize \begin{align}
    x_{t+1} = A x_{t} + B u_{t},
    \label{eq:system}
\end{align} \normalsize
with the fully measurable state $x_t \in \mathbb{R}^{{n_x}}$ and input $u_t \in \mathbb{R}^{{n_u}}$. A state-feedback policy $u_t = \pi(x_t)$ must stabilize the plant and minimize the infinite-horizon quadratic cost
\normalsize \begin{align}
    \mathcal{J}(x_0;\pi) := \sum_{t=0}^{\infty}  x_t^\top Q x_t + u_t^\top R u_t,
    \label{eq:cost}
\end{align} \normalsize
where $Q\succeq 0$ and $R\succ 0$. Under the standard well-posedness assumptions that (i) $(A,B)$ is stabilizable and (ii) $(Q^{1/2},A)$ is detectable, this is the infinite-horizon, discrete-time LQR problem~\cite{anderson1990optimal}, whose solution is $u_t^\star = \pi^\star(x_t) = -K^\star x_t$ with
\normalsize \begin{align}
    \label{eq:lqr_gain}
    K^\star = \bigl(R + B^\top P B\bigr)^{-1} B^\top P A,
\end{align} \normalsize
where $P$ solves the Algebraic Riccati Equation (ARE)
\normalsize \begin{align}
    P = Q + A^\top P A - A^\top P B \bigl(R + B^\top P B\bigr)^{-1} B^\top P A,
\label{ARE}
\end{align} \normalsize
and the optimal cost is $\mathcal{J}(x_0;\pi^\star) = x_0^\top P x_0$.

We are given a set of such LQR problems $\{ A^{(i)}, B^{(i)}, Q^{(i)}, R^{(i)} \}_{i=1}^N$, with possibly different state and input dimensions $(n_x^{(i)},n_u^{(i)})$ and optimal controllers $\pi^{\star,(i)}$, $i=1,\ldots,N$; the superscript refers to a specific problem instance. Our goal is a single universal policy $\pi_\phi\left(\{x_l\}_{l=t-w}^t,Q,R\right)$, parameterized by $\phi \in \R^k$, that is stabilizing and nearly optimal for each instance. The policy is static but uses the current and past $w$ state measurements ($w\in\mathbb{N}$ is the window length) together with the instance's cost matrices $(Q,R)$; including the cost as an input lets a single parameter vector produce cost-appropriate feedback across instances that share dynamics but differ in their objectives. Our design objective is small closed-loop sub-optimality over the $N$ instances with $J$ initial conditions each:
\normalsize
\normalsize \begin{align}
    \argmin_{\phi}\ \sum_{i=1}^N \sum_{j=1}^J \left[ \mathcal{J}^{(i)}\big(x_0^{(i,j)};\pi_\phi{\big)} - \mathcal{J}^{(i)}{\big(}x_0^{(i,j)}{;} \pi^{\star,(i)}{\big)} \right],
    \label{eq:primary_objective}
\end{align} \normalsize
\normalsize
where $\mathcal{J}^{(i)}$ is the LQR cost of problem $i$ as defined by $Q^{(i)}$ and $R^{(i)}$. A policy that makes \eqref{eq:primary_objective} small is our target. The trained policy also serves as an initializer for fine-tuning on unseen systems.

\section{Approach}
\label{sec:appraoch}
We learn a single universal control policy
$u_t=\pi_\phi\left(\{x_l\}_{l=t-w}^t,{Q,R}\right)$ in three steps:
a system-wise standardization maps the data $(u_t,\{x_l\}_{l=t-w}^t,{Q,R})$
to a standard form $(\bar{u}_t,S_t)$ (Section~\ref{subsec:generalizable_optimal_control});
a transformer realizes a policy $\bar{\pi}_\phi$ mapping $S_t$ to {a prediction $\hat{\bar{u}}_t$ of $\bar{u}_t$}
(Section~\ref{subsec:transformers}); and de-standardization recovers the physical $u_t$ (Section~\ref{subsec:training}). Section~\ref{subsec:finetuning}
describes fine-tuning for unseen systems.

\subsection{Data Collection and Processing}
\label{subsec:datacoll}\label{subsec:generalizable_optimal_control}

{Fix LQR problem $i\in\{1,\ldots,N\}$ with matrices $(A^{(i)}, B^{(i)}, Q^{(i)}, R^{(i)})$.} Simulating its optimal state feedback $\pi^{\star,(i)}(x_t) = -K^{\star,(i)} x_t$ from $J$ initial conditions $x_0^{(i,j)}$, $j=1,\ldots,J$, yields the {length-$T$ ($T\in\mathbb{N}$)} state and control sequences $X^{(i,j)} =\{x_{0}^{(i,j)}, \cdots, x_{T-1}^{(i,j)}\}$ and $U^{(i,j)}= \{u_{0}^{(i,j)}, \cdots, u_{T-1}^{(i,j)}\}$.

To ensure coverage and diversity, the $N$ instances span mechanical, electrical, and robotic LTI systems; each nominal system's parameters are varied to form a collection of variants, and the trajectories start from random initial conditions within a predefined bound.

Since an instance is the full quadruple $(A^{(i)},B^{(i)},Q^{(i)},R^{(i)})$, changing $Q^{(i)}$ or $R^{(i)}$ alone already yields a new instance with a different optimal gain $K^{\star,(i)}$. We therefore vary the cost together with the dynamics. Because the policy receives $(Q,R)$ as input, it learns to map a cost specification to its optimal feedback rather than to a single fixed objective. Objectives outside the training range can still be handled by the lightweight fine-tuning of Section~\ref{subsec:finetuning}.

We recast the system-specific LQR data into a sequence-to-action form through system-wise standardization, zero-padding, cost encoding, and masking.

\textbf{System-wise standardization.} To remove scale differences across systems, we standardize each system's data with its own statistics. Fix a problem $i$ and drop its superscript, writing $x_t^{(j)}$ and $u_t^{(j)}$ for the state and input of trajectory $j$ at time $t$. {Let $\mu_x,\sigma_x\in\R^{n_x}$ denote the entrywise mean and standard deviation of the states $x_t^{(j)}$ over all trajectories $j=1,\ldots,J$ and times $t=0,\ldots,T-1$, and let $\mu_u,\sigma_u\in\R^{n_u}$ denote the analogous statistics of the inputs. These statistics are computed once per system. They are pooled over the training trajectories of all problem instances of that system. Each sample is then standardized entrywise,
\begin{align}
\hat{x}_t^{(j)} &= (x_t^{(j)}-\mu_x)\oslash\sigma_x, \\
\hat{u}_t^{(j)} &= (u_t^{(j)}-\mu_u)\oslash\sigma_u,
\end{align}
where $\oslash$ denotes entrywise division {and $\sigma_x,\sigma_u$ are entrywise positive},} mapping all systems into a shared representation space. At inference the predicted control is de-standardized with the same statistics.

\textbf{Zero-padding.} To unify dimensions, let $\mathrm{Pad}(\cdot,\,n)$ append trailing zeros to bring its argument up to dimension $n$, and pad the {standardized} data to the maximum dimensions {$n_x^{\mathrm{max}}:=\max_i n_x^{(i)}$ and $n_u^{\mathrm{max}}:=\max_i n_u^{(i)}$} as
\begin{align}
\bar{x}_t^{(j)} = \mathrm{Pad}\big(\hat{x}_t^{(j)}, n_x^{\mathrm{max}}\big),
\end{align}
and
\begin{align}
\bar{u}_t^{(j)} = \mathrm{Pad}\big(\hat{u}_t^{(j)}, n_u^{\mathrm{max}}\big).
\end{align}
A {window of $w+1$ states} (unit stride; larger strides are possible) slides over the states, letting recent history act as implicit system identification: the input at time $t$ spans $\bar{x}_{t-w}^{(j)},\dots,\bar{x}_t^{(j)}$ and the target is the {current} control $\bar{u}_t^{(j)}$.

\textbf{Cost encoding.} To inform the transformer of the objective it imitates, we append an encoding of the cost. All cost matrices here are diagonal with positive entries,
$Q=\mathrm{diag}(q)$ and $R=\mathrm{diag}(r)$ with $q\in\R^{n_x}_{>0}$, $r\in\R^{n_u}_{>0}$, so $(q,r)$ specifies the objective exactly. We take their entrywise logarithm, which compresses the wide magnitude range across systems, and zero-pad,
\begin{align}
\bar{q} := \mathrm{Pad}\big(\log q,\, n_x^{\mathrm{max}}\big), \\
\bar{r} := \mathrm{Pad}\big(\log r,\, n_u^{\mathrm{max}}\big),
\end{align}
then attach $(\bar{q},\bar{r})$ to every padded state, giving the input
\normalsize \begin{align}
S_t^{(j)} = \begin{bmatrix} (\bar{x}_{t-w}^{(j)})^\top \oplus {\bar{q}^\top} \oplus {\bar{r}^\top} \\ \vdots\\ (\bar{x}_{t}^{(j)})^\top \oplus {\bar{q}^\top} \oplus {\bar{r}^\top} \end{bmatrix} \label{eq:S}
\end{align} \normalsize
of fixed size $(w+1)\times d_{\mathrm{in}}$, $d_{\mathrm{in}}=2n_x^{\mathrm{max}}+n_u^{\mathrm{max}}${, where $\oplus$ denotes vector concatenation}.

\textbf{Masking.} To avoid penalizing padded control entries, define the binary mask $\kappa := \mathrm{Pad}(1_{n_u},\, n_u^{\mathrm{max}})$ ($n_u$ ones {followed by} $n_u^{\mathrm{max}}-n_u$ zeros), where $1_n\in\R^n$ is the all-ones vector (for any dimension $n$). Each sample is the triplet
\begin{align}
(S_t^{(i,j)},\bar{u}_t^{(i,j)},\kappa^{(i)}),
\end{align}
and the dataset $\mathcal{D}$ collects these over all\footnote{For $t<w$ the missing past states are zero-padded to keep the length $w+1$.} $t\in\{0,\dots,{T-1}\}$, trajectories $j\in\{1,\dots,J\}$, and problems $i\in\{1,\dots,N\}$.

The input uses the state history and {the cost {$(\bar{q},\bar{r})$}} rather than past controls: since $u_t=-K^\star x_t$, the past controls are already determined by the state evolution pattern, and the closed-loop signature $x_{t+1}=(A-BK^\star)x_t$---its decay rate, oscillation, and transient response---already encodes their effect. Conditioning on past controls would also feed the policy its own outputs at inference, causing auto-regressive error accumulation.

\subsection{Transformers}
\label{subsec:transformers}

Attention transformers \cite{vaswani2017attention} are the state-of-the-art neural models {for} learning complex temporal patterns, especially in {large language models} \cite{chang2024survey}. This section first describes the single-transformer neural model. The extension to a multi-transformer neural model is then discussed.

First, we introduce the design of a single transformer block. Consider the sequence $S_t$ as defined in \eqref{eq:S}, where we drop the superscript $j$ for simplicity.
The transformer realizes a policy on this standardized data,  $\bar{\pi}_\phi:\mathbb{R}^{(w+1)\times d_{\mathrm{in}}}\to\mathbb{R}^{n_u^{{\mathrm{max}}}}$, with parameters $\phi$. This policy takes the input data $S_t$ at time $t$ and generates the {predicted} standardized control
{$\hat{\bar{u}}_t=\bar{\pi}_\phi(S_t)$}. The mapping from this {prediction $\hat{\bar{u}}_t \in \R^{n_u^{\mathrm{max}}}$} back to the actual control $u_t \in \R^{n_u}$ is discussed later.

The first transformer operation embeds the sequence into width $d_{\mathrm{m}}$ by a time-wise affine map applied in matrix form,
\begin{align}
\mathcal{Z} = S_tW_{\mathrm{in}}^{\top} + 1_{w+1}b_{\mathrm{in}}^{\top}\ \in\ \mathbb{R}^{(w+1)\times d_{\mathrm{m}}},
\end{align}
where $W_{\mathrm{in}}\in\mathbb{R}^{d_{\mathrm{m}}\times d_{\mathrm{in}}}$, and $b_{\mathrm{in}}\in\mathbb{R}^{d_{\mathrm{m}}}$. A learned positional matrix $\mathcal{P}\in\mathbb{R}^{(w+1)\times d_{\mathrm{m}}}$ is then added to encode temporal order,
\begin{align}
H = \mathcal{Z} + \mathcal{P} \ \in\ \mathbb{R}^{(w+1)\times d_{\mathrm{m}}}.
\label{eq:H}
\end{align}

Fix a head index {$p\in\{1,\dots,h\}$} with head width $d_{h}$ satisfying $d_{\mathrm{m}} = hd_{h}$. The per-head query, key, and value matrices are obtained from $H$ by linear projections,
\begin{align}
{\mathcal{Q}^{(p)}} = H{W_{\mathcal{Q}}^{(p)}},\quad
{\mathcal{K}^{(p)}} = H{W_{\mathcal{K}}^{(p)}},\quad
{\mathcal{V}^{(p)}} = H{W_{\mathcal{V}}^{(p)}},
\end{align}
where {$W_{\mathcal{Q}}^{(p)},W_{\mathcal{K}}^{(p)},W_{\mathcal{V}}^{(p)}\in\mathbb{R}^{d_{\mathrm{m}}\times d_{h}}$}
and {$\mathcal{Q}^{(p)},\mathcal{K}^{(p)},\mathcal{V}^{(p)}\in\mathbb{R}^{(w+1)\times d_{h}}$}. The {head-$p$ attention weights are obtained by applying \texttt{softmax} row-wise to the scaled logits,}
\begin{align}
{\mathcal{A}^{(p)} = \texttt{softmax}\Big(\tfrac{\mathcal{Q}^{(p)}(\mathcal{K}^{(p)})^{\top}}{\sqrt{d_{h}}}\Big)\ \in\ \mathbb{R}^{(w+1)\times (w+1)}.}
\end{align}

A single matrix multiplication computes the {head-$p$} context as
\begin{align}
{H^{(p)} = \mathcal{A}^{(p)}\mathcal{V}^{(p)}}\ \in\ \mathbb{R}^{(w+1)\times d_{h}},
\end{align}
which aggregates values according to the attention weights. Having formed {$H^{(p)}$} for all heads, {we assemble the multi-head output} by concatenation followed by an output projection,
\begin{align}
H^{\mathrm{cat}} &= \big[H^{(1)}, \cdots, H^{(h)}\big]\ \in \mathbb{R}^{(w+1)\times {d_{\mathrm{m}}}},\\
{\mathcal{M}} &= H^{\mathrm{cat}}W_{O} \in \mathbb{R}^{(w+1)\times d_{\mathrm{m}}},
\end{align}
with $W_{O}\in\mathbb{R}^{{d_{\mathrm{m}}}\times d_{\mathrm{m}}}$.

Next, we define the normalization layer $\texttt{LN}(x)$ as follows. Denote the mean and variance {of} a vector $x\in\mathbb{R}^{d}$ by {$E[x]$ and $\mathrm{Var}[x]$}. The layer normalization recenters and rescales $x$ and applies an affine transform:
\begin{align}
\texttt{LN}(x)
\;=\;
\gamma \odot \frac{x-{E[x]}1_d}{\sqrt{{\mathrm{Var}[x]}+\rho}}+\beta,
\end{align}
where $\gamma,\beta\in\mathbb{R}^{d}$ are learnable parameters. Here $\odot$ denotes element-wise multiplication, and $\rho$ is a small constant for numerical stability {(set to $10^{-5}$)}. $\texttt{LN}(\cdot)$ can also {be applied to a matrix row-wise}.

Residual addition and layer normalization stabilize the feature scale as follows{:}
\begin{align}
\mathcal{C} = \texttt{LN}\big(H + {\mathcal{M}}\big) \in \mathbb{R}^{(w+1)\times d_{\mathrm{m}}}.
\end{align}
This is then passed through a position-wise feed-forward mapping written in matrix form,
\begin{align}
F &= \texttt{GELU} \big(\mathcal{C}W_{1} + 1_{w+1}b_{1}^{\top}\big)\ \in\ \mathbb{R}^{(w+1)\times d_{\mathrm{ff}}},\\
G &= FW_{2} + 1_{w+1}b_{2}^{\top} \in \mathbb{R}^{(w+1)\times d_{\mathrm{m}}}, \label{eq:G}
\end{align}
where $W_{1}\in\mathbb{R}^{d_{\mathrm{m}}\times d_{\mathrm{ff}}}$, $W_{2}\in\mathbb{R}^{d_{\mathrm{ff}}\times d_{\mathrm{m}}}$, $b_{1}\in\mathbb{R}^{d_{\mathrm{ff}}}$, {and} $b_{2}\in\mathbb{R}^{d_{\mathrm{m}}}$. $\texttt{GELU}$ \cite{hendrycks2016gaussian} is the activation applied element-wise, and $d_{\mathrm{ff}}$ is a design hyperparameter.

After computing \eqref{eq:G}, {we obtain the single transformer block output} by a second residual addition and normalization,
\begin{align}
Y = \texttt{LN}\big(\mathcal{C} + G\big)\ \in\ \mathbb{R}^{(w+1)\times d_{\mathrm{m}}}.
\label{eq:Y}
\end{align}
This concludes the single transformer block design. The function $\mathrm{Block}\big(H\big)$ describes this process (i.e., {the operations mapping $H$ in \eqref{eq:H} to $Y$ in \eqref{eq:Y}}).

Next, we form a multi-transformer-block design. Starting from the single-block output $Y\in\mathbb{R}^{(w+1)\times d_{\mathrm{m}}}$, {we produce the control} from \emph{multiple} transformer blocks by aggregating the last-time representations from each block. Let $L\in\mathbb{N}$ denote the number of blocks. For each $\ell=1,\dots,L$, apply one transformer block {in parallel to the same} hidden sequence $H$ to obtain
\begin{align}
Y^{(\ell)} = \mathrm{Block}_{\ell}\big(H\big) \in \mathbb{R}^{(w+1)\times d_{\mathrm{m}}}.
\end{align}
We extract the last row from each block output $Y^{(\ell)}$ and convert it
to a column vector
\begin{align}
{y^{(\ell)}} =
\left( Y^{(\ell)}(w+1,:) \right)^{{\top}}
\in \mathbb{R}^{d_{\mathrm{m}}}.
\end{align}
Then, these vectors are stacked by feature-wise concatenation to form the layer-aggregated feature,
\begin{align}
\mathcal{R} = \big[{y^{(1)};y^{(2)};\dots;y^{(L)}}\big] \in \mathbb{R}^{Ld_{\mathrm{m}}}.
\end{align}
The aggregated feature is then mapped to the control space by a single linear readout,
\begin{align}
{\hat{\bar{u}}_t} = \bar{\pi}_\phi(S_t) =W_{\mathrm{out}}\mathcal{R} + b_{\mathrm{out}} \in \mathbb{R}^{n_u^{\mathrm{max}}},
\end{align}
with $W_{\mathrm{out}}\in\mathbb{R}^{n_u^{\mathrm{max}}\times (Ld_{\mathrm{m}})}$ and $b_{\mathrm{out}}\in\mathbb{R}^{n_u^{\mathrm{max}}}$.

In summary, the sequence $H$ yields $Y^{(\ell)}$, which yields {$y^{(\ell)}$}; concatenating {$\{y^{(\ell)}\}_{\ell=1}^{L}$} gives $\mathcal{R}$, and finally $\mathcal{R}$ determines {$\hat{\bar{u}}_t$}. All weight matrices, bias vectors, positional embeddings{, input embedding}, attention projections, feed-forward mappings, normalization layers, and the final readout collectively constitute $\phi$.
\subsection{Training and Inference}
\label{subsec:training}
Training minimizes a \emph{masked Cauchy} loss, which only evaluates error on valid control dimensions, over {randomly sampled mini-batches $\mathcal{B}\subset\mathcal{D}$}:
\normalsize \begin{align}
\mathcal{L}
= \frac{1}{|\mathcal{B}|} \sum_{j \in \mathcal{B}}
\ln \left( 1 + \left\| \frac{\kappa^{(j)}\odot(
\bar{\pi}_\phi(S^{(j)}) - \bar{u}^{(j)})}{\xi}\right\|_{{2}}^2 \right),
\end{align} \normalsize
where $\xi$ is a scalar parameter, and $j$ refers to a specific sample in this batch. The parameters are updated with AdamW~\cite{loshchilov2018decoupled}. This supervised loss is a surrogate for the design objective \eqref{eq:primary_objective}. Upon completion of training, inference on a specific problem $i$ can be achieved by de-standardizing a predicted control {$\hat{\bar{u}}_t^{(i)}$} associated with problem $i$, and extracting the first $n_u^{(i)}$ entries. This results in $u_t^{(i)}$, which serves as the final prediction by the transformer. Specifically,
\begin{align}
u_t^{(i)} = \sigma_u^{(i)}\odot{\hat{\bar{u}}_t^{(i)}}(1:n_u^{(i)}) + \mu_u^{(i)},
\end{align}
with $\odot$ applied entrywise.

\subsection{Fine-Tuning for Unseen Systems}
\label{subsec:finetuning}
For systems not seen during training, minimal few-shot fine-tuning---updating the weights on a few trajectories from the unseen system family---adapts the trained transformer with little retraining. The standardization statistics of an unseen system are computed from these fine-tuning trajectories.

\section{Results}
\label{sec:results}

\subsection{Setup} Training data are generated from the seen systems by perturbing the nominal dynamics {entrywise} by up to $\delta=\pm10\%$, pairing every perturbed instance with every cost setting, solving \eqref{ARE} for the gain $K^{\star,(i)}$ of each resulting instance $(A^{(i)},B^{(i)},Q^{(i)},R^{(i)})$, and rolling out optimal trajectories from random initial conditions until $1.5\times10^{6}$ sequences are collected per system ($95\%/5\%$ train/test split). We evaluate on the $28$ discrete-time MIMO LTI systems of \autoref{tab:lti_systems}. For each seen system we form nine cost settings: the nominal $(Q,R)$ plus mild log-uniform perturbations of their diagonals (up to a factor of $3$ per entry). Unseen systems receive minimal fine-tuning (two epochs on their data) before evaluation. \autoref{tab:hyperparameters} lists all hyperparameters.

\begin{table}[h]
\normalsize
\centering
\caption{Benchmark systems}
\label{tab:lti_systems}
\begin{tabular}{@{}>{\raggedright\arraybackslash}p{0.98\linewidth}@{}}
\toprule
\textbf{Seen Systems:}
\textbf{1.}~Simple Pendulum, \textbf{2.}~Two Link Arm, \textbf{3.}~Spring Damper,
\textbf{4.}~Suspension System, \textbf{5.}~DC Motor,  \textbf{6.}~Three Link Manipulator,
\textbf{7.}~Differential Drive Robot,
\textbf{8.}~SCARA Robot, \textbf{9.}~Omnidirectional Robot,
\textbf{10.}~Cable Driven Robot, \textbf{11.}~Flexible Joint Robot,  \textbf{12{.}}~Six DOF Manipulator,
\textbf{13.}~Dual Arm Robot, \textbf{14.}~Double Integrator, \textbf{15.}~Lotka Volterra.\\
\midrule
\textbf{Unseen Systems:}
\textbf{16.}~Inverted Pendulum, \textbf{17.}~Segway Robot, \textbf{18.}~Asymmetric Oscillator,
\textbf{19.}~Active Mass Damper, \textbf{20.}~Coupled Oscillators, \textbf{21{.}}~Damped Oscillator,
\textbf{22.}~Triple Mass Spring, \textbf{23.}~Electromechanical Actuator, \textbf{24{.}}~Thermal System,
\textbf{25.}~Fluid Tank, \\ \textbf{26.}~Vibrating Beam, \textbf{27.}~Motor Generator,
\textbf{28.}~Mechanical Linkage.\\
\bottomrule
\end{tabular}
\end{table}

\begin{table}[h]
\normalsize
\centering
\caption{Hyperparameters}
\label{tab:hyperparameters}
\begin{tabular}{@{}>{\raggedright\arraybackslash}p{0.98\linewidth}@{}}
\toprule
\textbf{Data:} \\ \mbox{time step $\Delta t{=}0.02$\,s}, \mbox{rollout length $T{=}1250$}, \mbox{$1.5{\times}10^{6}$ sequences/system}, \mbox{perturbation $\delta{=}{\pm}10\%$}.\\
\midrule
\textbf{Architecture:} \\ \mbox{window $w{=}12$}, \mbox{max.\ dimensions $n_x^{\mathrm{max}}{=}12$}, \mbox{$n_u^{\mathrm{max}}{=}6$}, \mbox{embedding width $d_{\mathrm{m}}{=}64$}, \mbox{heads $h{=}16$}, \mbox{blocks $L{=}4$}, \mbox{feed-forward width $d_{\mathrm{ff}}{=}256$}.\\
\midrule
\textbf{Training:} \\ \mbox{epochs $N_{\mathrm{epochs}}{=}500$}, \mbox{batch size $|\mathcal{B}|{=}4096$}, \mbox{learning rate $\eta{=}5{\times}10^{-5}$}, \mbox{Cauchy scale $\xi{=}1$}.\\
\bottomrule
\end{tabular}
\end{table}

\subsection{Metric} For system $i$ and initial state $x_0$, the \emph{relative sub-optimality} of a closed-loop rollout is
\normalsize \begin{align}
\Delta^{(i)}(x_0) := \frac{\mathcal{J}^{(i)}(x_0;\pi_\phi) - \mathcal{J}^{(i)}(x_0;\pi^{\star,(i)})}{\mathcal{J}^{(i)}(x_0;\pi^{\star,(i)})}.
\end{align} \normalsize
Each system is evaluated on-policy over perturbed variants (ten for seen systems, five for unseen; $\delta=\pm10\%$), the nine training cost settings, and five initial conditions per pair, with the transformer conditioned on each pair's $(Q,R)$---$450$ ($225$) rollouts per seen (unseen) system.

\subsection{Family-Wide Performance}
\begin{figure}[h]
    \centering
    \includegraphics[width=\linewidth]{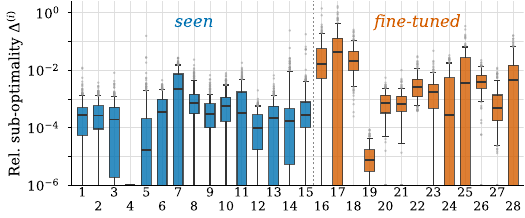}
    \caption{Relative sub-optimality $\Delta^{(i)}$ for each system in \autoref{tab:lti_systems}.}
    \label{fig:relative}
\end{figure}

\autoref{fig:relative} shows the distribution of $\Delta^{(i)}$ for every system. No unstable rollout was observed across the $9{,}675$ evaluated trajectories. On seen systems the median sub-optimality is $0.022\%$ (95th percentile $0.38\%$): the closed-loop cost is within a few hundredths of a percent of optimal. Fine-tuned unseen systems follow closely, with a median of $0.19\%$ (95th percentile $8.3\%$); the similarity of the two distributions indicates that the learned mapping generalizes across systems rather than memorizing {them}. The largest gaps occur on the Segway, asymmetric oscillator, and inverted pendulum (medians $4.4\%$, $2.1\%$, and $1.6\%$), where fast unstable or lightly damped modes amplify small imitation errors; even there, all evaluated rollouts remained stable.

\subsection{Closed-Loop Trajectories}
\begin{figure}[h]
    \centering
    \includegraphics[width=\linewidth]{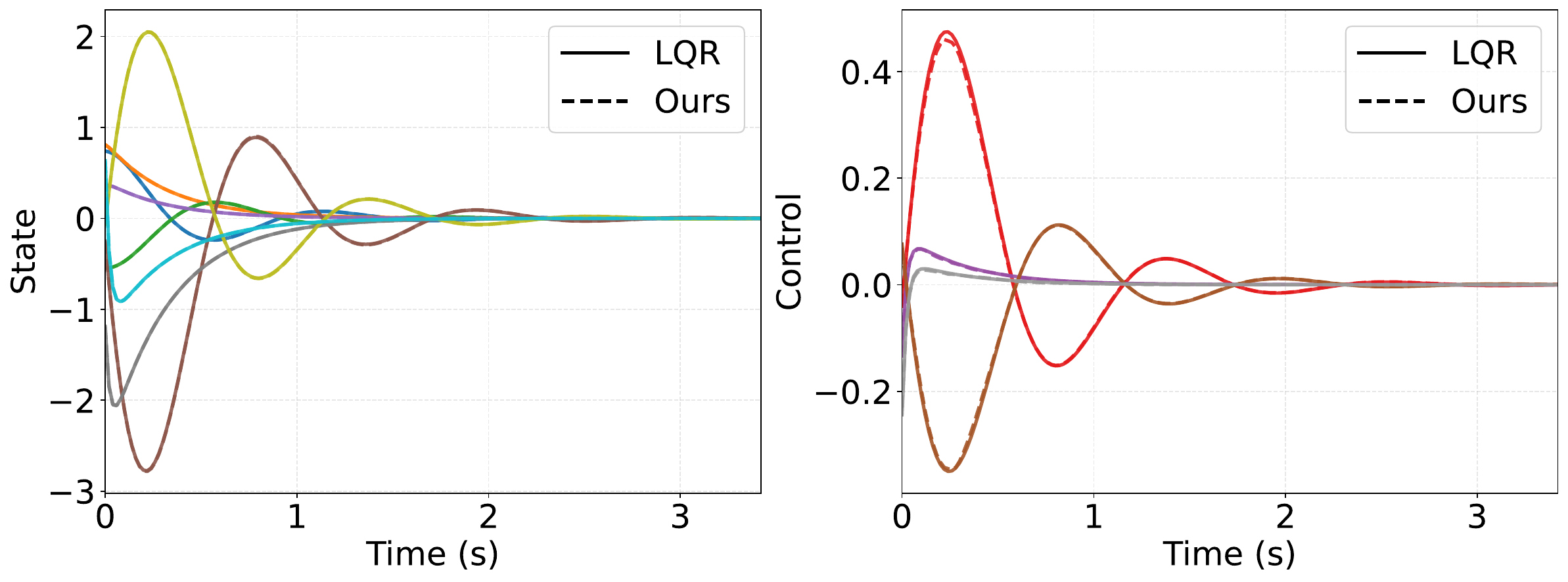}
    \caption{Performance on the dual arm robot (system \#13, seen).}
    \label{fig:DualArm}
\end{figure}

\begin{figure}[h]
    \centering
    \includegraphics[width=\linewidth]{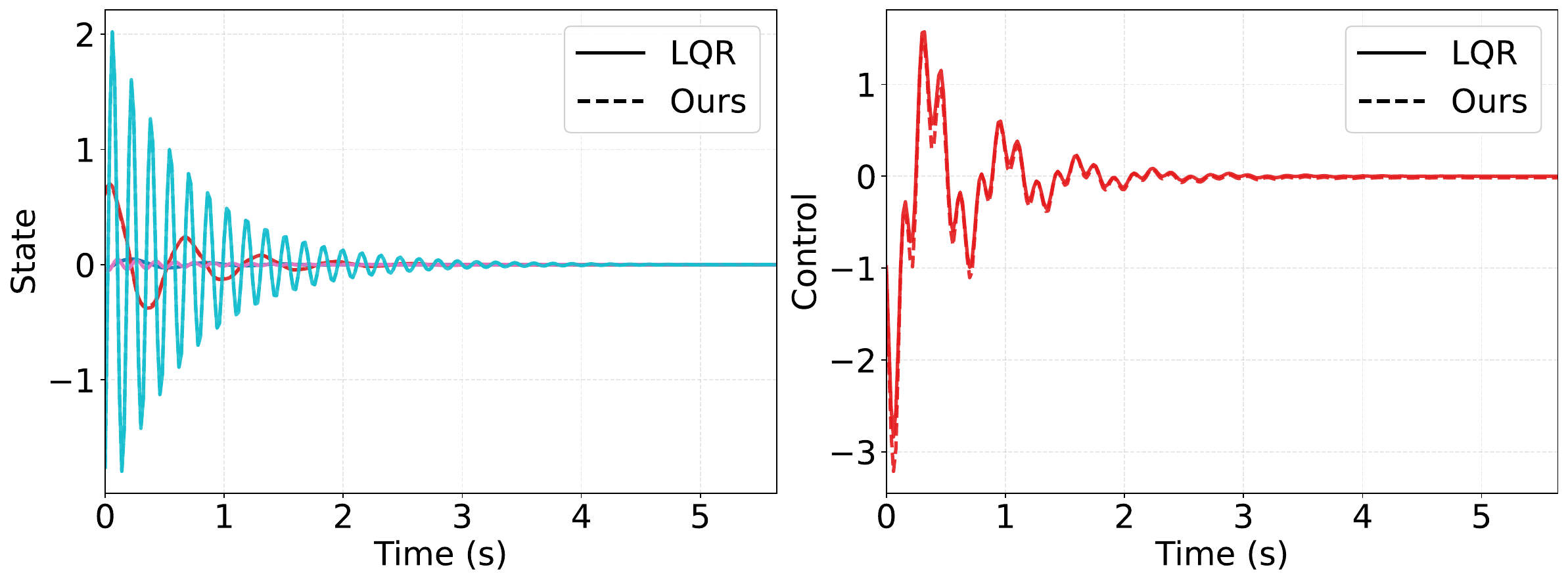}
    \caption{Performance on the vibrating beam (system \#26, unseen).}
    \label{fig:VibratingBeam}
\end{figure}

\autoref{fig:DualArm} and \autoref{fig:VibratingBeam} show representative rollouts (states left, control right). On the seen {dual arm robot}, $\pi_\phi$ reproduces the optimal control command and matches its transient almost exactly; on the unseen vibrating beam, the fine-tuned policy is likewise nearly indistinguishable from the optimal controller.

\subsection{Ablation} 
\begin{figure}[h]
    \centering
    \includegraphics[width=\linewidth]{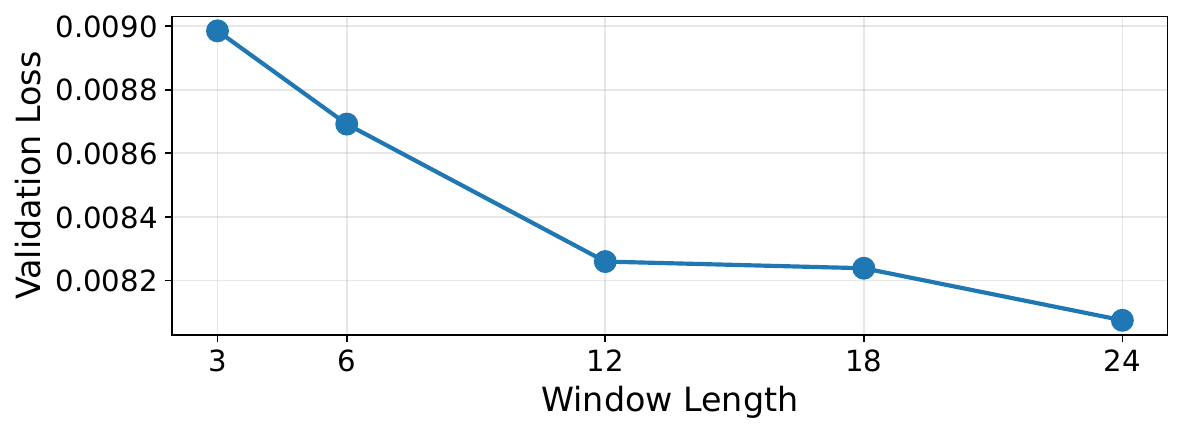}
    \caption{Validation loss versus window length $w$.}
    \label{fig:Ablation}
\end{figure}

\autoref{fig:Ablation} varies the window length $w$ and reports the corresponding validation loss. It decreases with $w$ and improves marginally beyond $w=12$. This is consistent with the window serving as implicit system identification. Thus,  we adopt $w=12$ as the accuracy--computation trade-off.

\section{Conclusions}
\label{sec:conclusions}
We presented a single transformer policy that imitates optimal LQR feedback across a family of heterogeneous LTI systems. Rather than a plant model, the policy uses a window of recent states together with the cost matrices. System-wise standardization, zero-padding, and masking give it a shared representation that spans different state and input dimensions, and the cost encoding lets one set of weights serve different objectives. Across $28$ systems and {$9{,}675$} rollouts, no unstable rollout was observed. It reaches a median relative sub-optimality of $0.022\%$ on seen systems and $0.19\%$ on unseen systems after lightweight fine-tuning. A transformer can therefore act as a generalizable near-optimal controller for a structured family of linear systems.

Several limitations frame the next steps. The results are demonstrated empirically rather than certified, the plants are noise-free, and unseen systems still require fine-tuning rather than zero-shot transfer. We will pursue stability and sub-optimality guarantees for the learned policy, study its behavior under process and measurement noise{, compare it against explicit identify-then-design methods,} and work toward zero-shot transfer to unseen systems. Extending the cost encoding beyond diagonal objectives and scaling to larger dimensions are natural further directions.

\section*{Acknowledgment}
T. Bin Mohaya and M. F. AL-Sunni acknowledge financial support from King Fahd University of Petroleum and Minerals (KFUPM), Dhahran, Saudi Arabia.


\bibliographystyle{IEEEtran}
\bibliography{references}
\end{document}